\documentclass[epsfig,rotate]{elsart}

\usepackage{graphicx}

\usepackage{amssymb}

\begin{document}

\begin{frontmatter}



\title{Calculation of the separation streamlines of barchans and transverse dunes}

\author[stuttgart,fortaleza]{H. J. Herrmann},
\author[fortaleza]{J. S. Andrade Jr.},
\author[stuttgart]{V. Schatz}, 
\author[stuttgart]{G. Sauermann} and
\author[stuttgart]{E. J. R. Parteli}
\address[stuttgart]{ICP, University of Stuttgart, Pfaffenwaldring 27, 70569, Stuttgart, Germany}
\address[fortaleza]{Departamento de F\'{\i}sica, Universidade Federal do Cear\'a, 60455-970, Fortaleza, CE, Brazil}

\address{}

\begin{abstract}

We use FLUENT to calculate the wind profile over barchans and transverse dunes. The form of the streamlines of flow separation at the lee side of the dunes is determined for a symmetric barchan dune in three dimensions, and for the height profile of a measured transverse dune field in the Len\c{c}\'ois Maranhenses.

\end{abstract}


\end{frontmatter}

\section{Introduction}
\label{introduction}

Sand dunes are one of the most beautiful patterns formed by the wind. The most well-known type of dune is the {\it{barchan}}, which arises under uni-directional wind and when there is not much sand on the ground \cite{bagnold}. This dune has a stoss-side, where sand is transported and deposited by the wind, and a steep slip face where avalanches take place maintaining dune stability and motion. {\it{Transverse dunes}} arise when sand availability increases and several barchans touch at their horns building a dune chain that propagates in the direction of the wind. The brink of the dune represents the beginning of the slip-face and introduces a sharp flow discontinuity. While at the stoss-side sand is transported downwind, at the lee side a recirculating flow is observed. The region of recirculating flow after the dune is known as {\it{separation bubble}} \cite{gerd_thesis}, and extends from the brink of the dune up to the reattachment point on the floor at a given distance from the brink. This distance has been found to depend on the dune height as well as on the dune shape at the brink \cite{lencois_2003}. Although the main processes of dune morphology, aeolian sand transport and avalanches, are rather well understood \cite{bagnold,gerd_thesis,tsoar_book}, very little is known about the separation bubble. Since a closed analytical solution that consistently describes the flow over the entire dune is hard to obtain, we need to measure the wind velocity and direction in the field to calculate the separation bubble and to introduce it into a phenomenological dune model. A first quantitative estimative of the wind flow can be obtained from the numerical calculation of the Navier Stokes equations that govern the behavior of the wind. Many programs, packages, and libraries have been developed in order to solve the Navier Stokes equations in different geometries and boundary conditions. We have chosen FLUENT V5.0 \cite{fluent} to calculate the flow over a symmetric barchan dune in three dimensions and for the measured height profile of a transverse dune field in the Len\c{c}\'ois Maranhenses, northeastern Brazil \cite{lencois_2003}.  In the next section we give a short description of the physical elements that define the boundary conditions of the problem. Our results are then presented and discussed in section 3. Conclusions are made in section 4.

\section{Wind profile and turbulent flow over a dune}

Sand transport takes place close to the ground, in the atmospheric boundary layer \cite{tsoar_book}, where even small wind speeds are associated with large Reynolds numbers. Thus, the flow at which aeolian sand transport occurs is always turbulent \cite{tsoar_book}. The velocity of the wind in the atmospheric boundary layer increases logarithmically with the height $z$:
\begin{equation}
v(z) = {\frac{u_{\ast}}{\kappa}}{\ln{\frac{z}{z_0}}}, \label{eq:logprofile}
\end{equation}
where $z_0$ denotes the roughness length of the surface, $\kappa=0.4$ is the von K\'arm\'an constant, and $u_{\ast} = {\sqrt{{\tau}/{\rho}}}$ is the shear velocity, where $\tau$ is the shear stress of the wind, and $\rho=1.225$kg m$^{-3}$ is the air density. This relation defines the boundary condition for the wind profile in the calculation of the turbulence model. 
Since the fluctuations of the turbulent flow can be of small scale and high frequency, the Navier Stokes equations are usually time-averaged, ensemble-averaged or otherwise manipulated to remove the small scale dynamics \cite{hinze}, which results in a modified set of equations that are computationally less expensive to solve. However, since the averaging can not literally be performed, these effective equations are not closed and they contain additional unknown terms. These terms have to be related to the averaged variables by a closure relation, called {\it{turbulence model}}, which may consist of two common methods: either {\em{Reynolds averaging}} (e.g. the Spalart-Allmaras model, different versions of the $k$-${\epsilon}$ model, and the Reynolds stress model) or {\em{filtering}} (e.g. large eddy simulation). We use the $k$-${\epsilon}$ model which calculates the turbulent kinetic energy $k$ and its rate of dissipation $\epsilon$ assuming that the flow is fully turbulent, and the effects of molecular viscosity are negligible. In the next section we present the results obtained with the FLUENT using the $k$-${\epsilon}$ model to calculate the wind flow over a barchan dune and a transverse dune field.

\section{Results and Discussion}
\label{results}

The first step is to generate a grid that defines the discretization of the volume and the boundaries. Here, a compromise between the number of cells and precision has to be found. A standard method to reduce the number of cells is to take unstructured grids that use smaller cells in areas where large gradients of the internal variables, velocity, pressure, turbulent energy, and turbulent dissipation are expected. The wake region after the sharp brink has been refined. Furthermore, the height of the cells has been increased exponentially with the distance from the ground. The wind is blowing in the positive $x$ direction, and the height is ploted on the $z$ axis. The boundaries are periodic in the $y-$direction. 

At the influx boundary, the velocity is fixed by the logarithmic velocity profile of Eq.(\ref{eq:logprofile}) using a shear velocity of $u_{\ast}=0.36$m$/$s, which was found to fit the field data of the measurements on a barchan dune in Jericoacoara \cite{jeri} and on the transverse dunes in the Len\c{c}\'ois Maranhenses \cite{lencois_2003}. At the opposite outflux boundary, the derivative of the velocity normal to the boundary is set to zero, which lets the fluid leave the simulation area. The bottom with the profile of the dunes is a fixed wall to which a roughness length of $z_0 = 2.5 \times 10^{-3}$m is associated \cite{gerd_thesis,lencois_2003}. Finally, the top of the simulation area at a height of 120m is implemented as a wall moving with the velocity of the outer flow which is defined by the velocity of the undisturbed logarithmic profile at the inflow. The calculation of the stationary wind field is started by initializing the variables (e.g. the velocity to zero). Then the equations for the velocity, the pressure, the turbulent energy, and the turbulent dissipation are iteratively calculated. If this process converges the solution is attained after a certain number of steps. 

\begin{figure}
\begin{center}
\includegraphics[width=0.70\columnwidth]{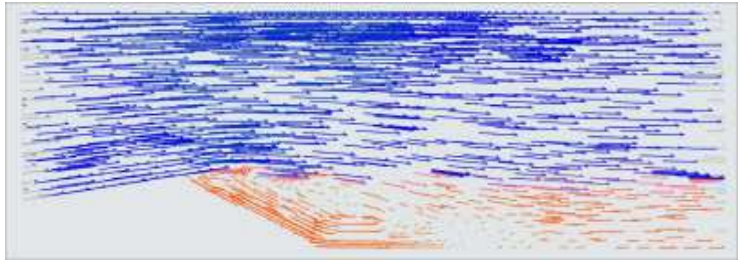}
\caption{Cut along the symmetry plane, the central slice of the dune. The depicted velocity vectors clearly show the separation of flow at the brink and a large eddy that forms in the wake of the dune \cite{gerd_thesis}.}
\label{fig:velocity_cut_black}
\end{center}
\end{figure}

\begin{figure}
\begin{center}
\includegraphics[width=0.5\columnwidth]{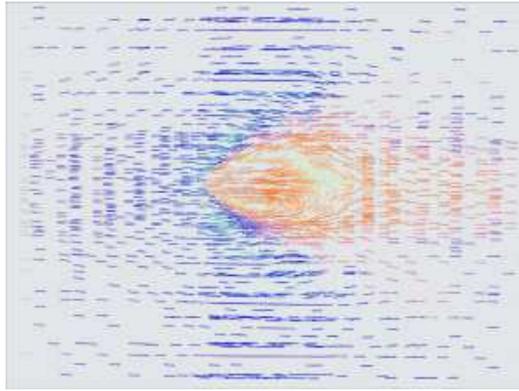}
\caption{Projection of the dune on the $x-y$ plane and the depicted velocity vectors reveal the three-dimensional structure of the large wake eddy \cite{gerd_thesis}.}
\label{fig:velocity_top_black}
\end{center}
\end{figure}

\begin{figure}
\begin{center}
\includegraphics[width=0.60\columnwidth]{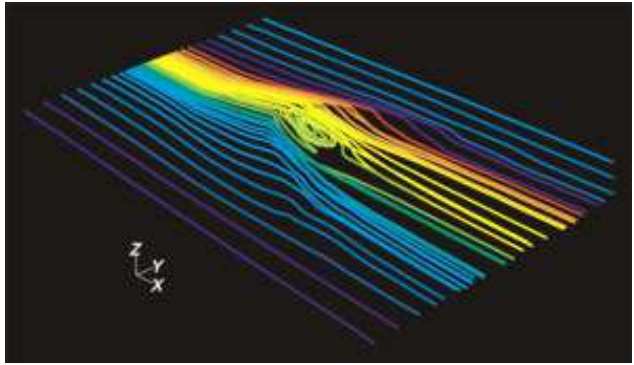}
\caption{The calculated lines are the trajectories of massless particles that are advected by the flow. It can be seen that the lateral component of the flow is very small.}
\label{fig:pathlines_black}
\end{center}
\end{figure}

Such a solution of the wind field is shown in Figures \ref{fig:velocity_cut_black} and \ref{fig:velocity_top_black} for a symmetric barchan in three dimensions. As we can see from these figures, a large eddy has been formed in the wake after the brink where the separation of the flow takes place, as expected due to the discontinuity of the brink. 
Furthermore, the lateral component of the wind velocity on the windward side of the dune is very small compared to the $x-$component. This is visualized by the streamlines depicted in Figure \ref{fig:pathlines_black}. Streamlines are the trajectories of massless particles which get advected with the flow. 

\begin{figure}
\begin{center}
\includegraphics[width=1.00\columnwidth]{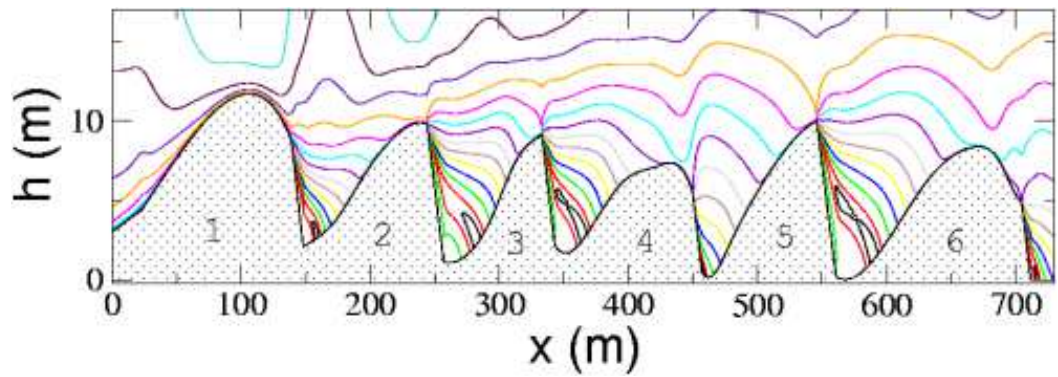}
\caption{Iso-velocity lines of the wind on the measured transverse dune profile calculated with FLUENT. Lines are shown in steps of $1.0$m/s from the bottom to the top.}
\label{fig:iso-velocity}
\end{center}
\end{figure}

\begin{figure}
\begin{center}
\includegraphics[width=0.95\columnwidth]{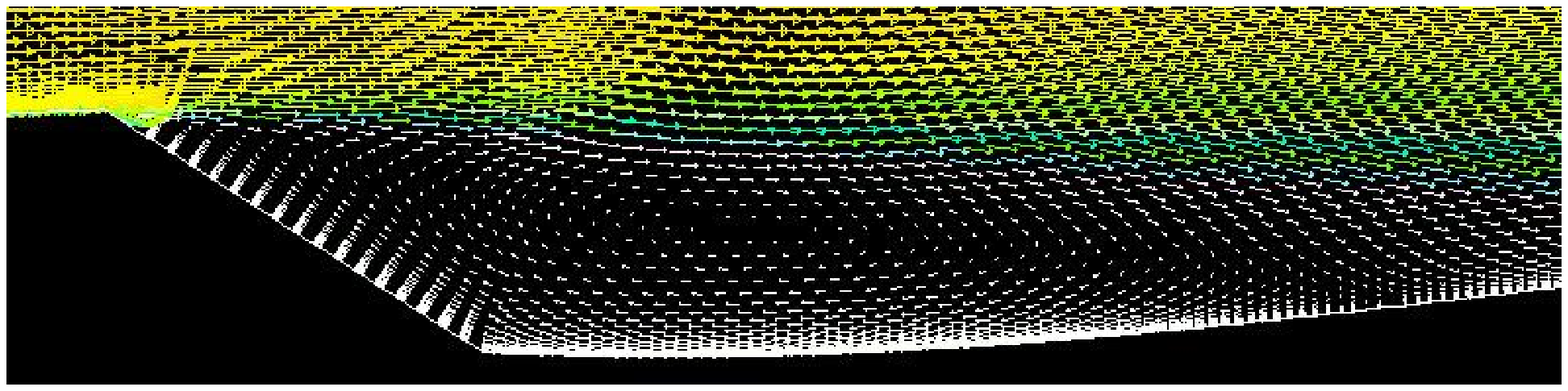}
\includegraphics[width=0.47\columnwidth]{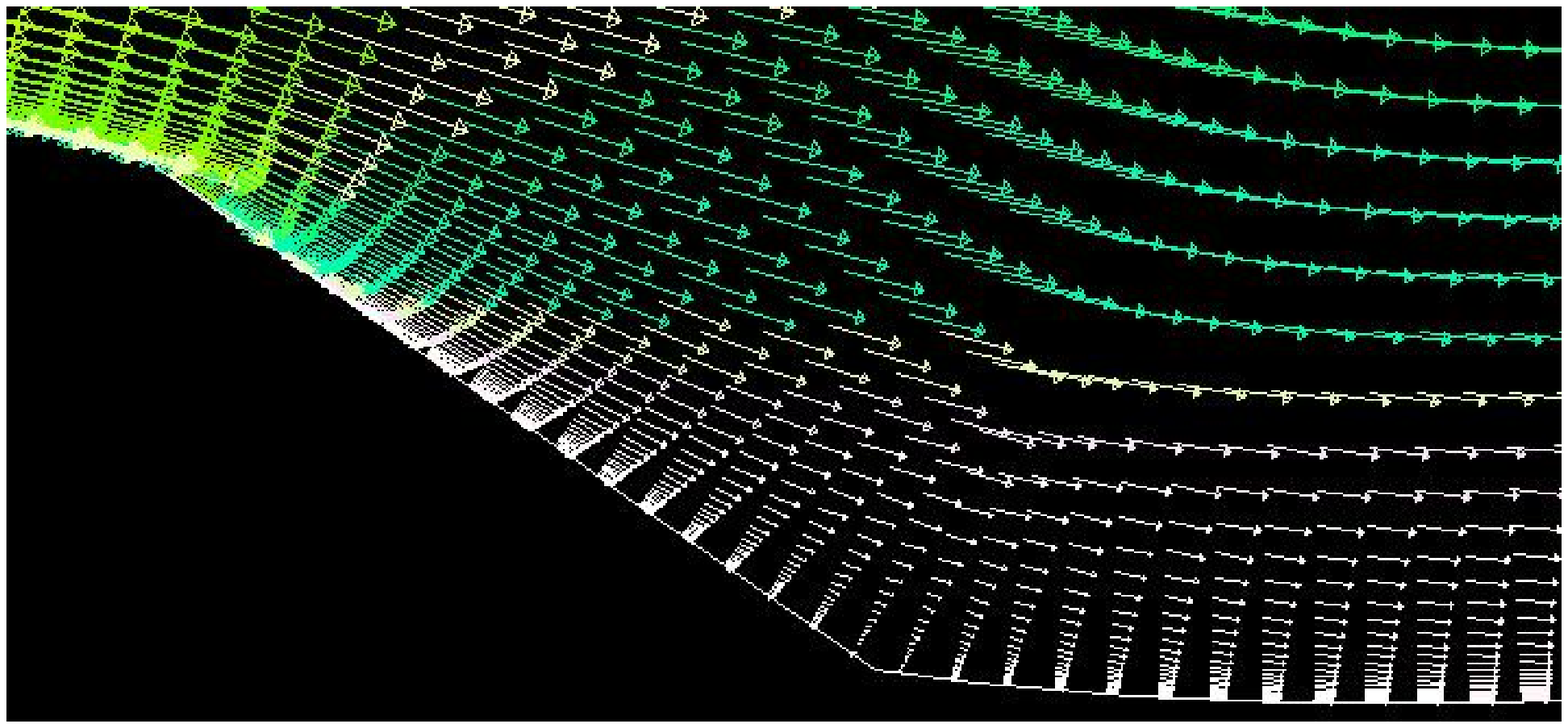}
\caption{Velocity vectors at the lee side of dunes number 4 (bottom) and 5 (top) of fig. \ref{fig:iso-velocity}.  The recirculating flow on the foot of the slip face denotes the separation bubble. It is large for dune number 5, but nearly non-existent for dune number 4.}
\label{fig:vectors_transverse}
\end{center}
\end{figure}

We next use FLUENT to study the wind over the two dimensional height profile of
a transverse dune field recently measured in the Len\c{c}\'ois Maranhenses
\cite{lencois_2003}. Figure \ref{fig:iso-velocity} shows the iso-velocity lines
calculated on this field. Wind velocity increases from the bottom to the top of
the figure in steps of $1.0$m/s. In fig. \ref{fig:vectors_transverse} we show, from the bottom to the top, 
the velocity vectors for the fourth and the fifth dune in the field, from the
left to the right in the profile.  We can see clearly that in the lee of the
rounded fourth dune, there is next to no recirculation, while the fifth dune
has a sharp brink and a large bubble.  In general, the length of the separation
bubble of transverse dunes with similar heights tends to be shorter for dunes
where the distance between the crest and the brink is larger. The very small
separation bubble of dune number four is a consequence of the fact that the
flow does not separate at the brink.  It is well known that flow separation
does not happen if a ``corner'' like the slip face brink forms an very obtuse
angle.  A detailed study of the control of the separation by the dune shape is
now being carried out.

The separation bubble of most of the dunes is larger than that used in ref.
\cite{lencois_2003} for a simulation of the creation of this dune field.  There
are two possible explanations.  One is that the dune field is not stable but
constantly changing because the wind only moves the part of the dunes outside
the separation bubble.  Another possibility is that gusts of turbulence which
occur from time to time move sand also within the separation bubble.  No
measurement of the wind speed at the foot of the dunes was carried out in
\cite{lencois_2003}, so at this time we cannot know which is the correct
explanation.

We refrain from a comparison between the flow over barchans and over transverse
dunes.  There were differences between the way the two simulations were
performed which may have influenced the results and make a comparison
inappropriate.  Notably the barchan simulation had to be done with an
inhomogeneous and in places significantly coarser grid than the simulation of
transverse dunes.  This was necessary because 3-dimensional simulations are
computationally expensive, but it means the simulations should not be directly
compared.

\vspace{-1cm}

\section{Conclusions}
\label{conclusions}

We have shown the wind velocity vectors and streamlines calculated for a tri-dimensional barchan and for measured transverse dunes in two dimensions using FLUENT. The recirculating flow at the lee side of the dunes defines the separation bubble. The results found here will be used to model the dynamics of barchans and transverse dunes.

\begin{ack}
We acknowledge O. Duran for helpful comments. This work was supported in part by the Max-Planck Price awarded to H. J. Herrmann (2002). E. J. R. Parteli acknowledges support from CAPES - Bras\'{\i}lia/Brazil.
\end{ack}

\end{document}